\documentclass{article}

\PassOptionsToPackage{numbers, compress}{natbib}

\usepackage[dblblindworkshop, final]{neurips_2026}
\workshoptitle{ML4Molecules}

\usepackage[utf8]{inputenc} 
\usepackage[T1]{fontenc}    
\usepackage{hyperref}       
\usepackage{url}            
\usepackage{booktabs}       
\usepackage{amsfonts}       
\usepackage{nicefrac}       
\usepackage{microtype}      
\usepackage{xcolor}         
\usepackage{graphicx}
\usepackage{amsmath}
\usepackage{xcolor}
\usepackage{siunitx}
\usepackage{comment}
\usepackage{wrapfig}
\usepackage{amssymb}
\usepackage{tabularx}
\usepackage{multirow}
\hypersetup{colorlinks=true,
    linkcolor = [HTML]{000000},
    citecolor = [HTML]{4B7081},
        urlcolor  = [HTML]{000000} }

\newcommand{\jn}[1]{{\color{purple}#1}} 
 
\newcommand{\bg}[1]{{\color{blue}#1}}
\newcommand{\bxi}{\boldsymbol{\xi}}
\newcommand{\bx}{\mathbf{x}}
\newcommand{\gFF}{\texttt{grappa-fullFF} }
\newcommand{\gBOND}{\texttt{grappa-bonded} }

\title{Learning consistent molecular mechanics\\force fields from first principles}

\author{%
  Berkay G\"unes \textsuperscript{1,2}, 
  Leif Seute \textsuperscript{1,3}, 
  Jigyasa Nigam \textsuperscript{4}, 
  Frauke Gr\"ater \textsuperscript{1,2,3} \\ \\
  \textsuperscript{1} Max Planck Institute for Polymer Research Mainz, Germany\\
  \textsuperscript{2} Heidelberg University, Germany\\
  \textsuperscript{3} Heidelberg Institute for Theoretical Studies, Germany\\
  \textsuperscript{4} Initiative for Computational Catalysis, Flatiron Institute, United States\\
  \texttt{\{guenesb, seutel, graeter\}@mpip-mainz.mpg.de, jnigam@flatironinstitute.org}
}

\begin{document}
\maketitle

\begin{abstract}
Classical force fields (FFs) remain the workhorse for large-scale simulations even as machine-learned interatomic potentials (MLIPs) approach ab initio accuracy. They decompose total configuration energies into simple effective interactions whose parameters are traditionally assigned based on atom or bond types, enabling efficient simulations but also limiting their ability to adapt across configurations. Recent machine learning approaches have improved the accuracy and transferability of bonded parameters in these FFs by inferring them as functions of local atomic environments, but still rely on empirical nonbonded parameters for practical simulations. In this work, we introduce a unified approach,  \texttt{grappa-fullFF}, which learns both bonded and nonbonded parameters \emph{consistently} and simultaneously from ab initio reference data.  
By incorporating physically inspired regularization via supervision of the electrostatic potential and an architecture that facilitates charge equilibration, our model recovers accurate electric response properties, achieves state-of-the-art accuracy on geometry optimization benchmarks, and reproduces the conformational sampling of both classical and existing machine-learned FFs, without relying on externally assigned nonbonded parameters. 
\end{abstract}

\section{Introduction}
Simulating molecular systems containing millions of atoms on nanosecond to millisecond timescales underlies numerous applications ranging from drug discovery and biomedical engineering to materials science~\cite{shaw2010atomic, borhani2012future, shibuta2017heterogeneity, morawietz2021machine, noe2020machine}. Such simulations require force fields that are both computationally efficient and sufficiently accurate to describe the potential energy and forces throughout the dynamics. Recent advances in atomistic modeling and deep learning have enabled MLIPs that predict energies and forces directly from three-dimensional molecular configurations with near ab initio accuracy \citep{bartok2010gaussian, smith2017ani, batatia2023macehigherorderequivariant, batzner2023nequip, mazitov2025pet, lysogorskiy2026graph}. Although MLIPs are orders of magnitude faster than quantum mechanical (QM) calculations, their computational cost remains prohibitive for such large-scale systems and long timescales~\cite{Unke21, wang2025design}, and classical molecular mechanics (MM) continues to be widely used to drive these simulations due to its exceptional efficiency \citep{morado2023does, 10.1021/ct9000685, 10.1371/journal.pcbi.1005659, 10.1021/ct400314y}. In MM, the Born-Oppenheimer potential energy of an atomic configuration $\mathbf{x}$ is approximated as a sum of physical contributions,
\begin{align}
    E_{\text{MM}}(\mathbf{x}, \Phi_{\text{FF}}) & 
    = E_\text{bonded}(\mathbf{x}, \Phi_{\text{bonded}}) + E_\text{nonbonded}(\mathbf{x}, \Phi_{\text{nonbonded}})
    \label{eq:mm_functional}
\end{align}
where $\Phi_{\text{FF}} = \{\Phi_{\text{bonded}}, \Phi_{\text{nonbonded}}\}$ represents the set of force field parameters.  $E_\text{bonded}(\mathbf{x}, \Phi_{\text{bonded}}) = $ $U_\text{bonds} + U_\text{angles} + U_\text{dihedrals}$ 
constitute \emph{bonded} interactions, while $E_\text{nonbonded}(\mathbf{x}, \Phi_{\text{nonbonded}}) = U_\text{vdW} + U_\text{Coulomb}$ are the \emph{nonbonded} interactions~\cite{Hagler2019, DBLP:journals/jcamd/Hagler19} comprising van der Waals (vdW) (typically the 12-6 Lennard-Jones~\citep{723ed04a-9dcc-30ee-832b-e24d0998621e}) and long-range Coulomb interactions. Traditionally, the parameters of these effective potentials have been refined empirically and tabulated for different classes of molecules, which restricts their accuracy and transferability. 
Recently, graph-based machine-learning frameworks~\cite{espaloma2022, 10.1039/d4sc00690a, grappa2025, 10.1021/acs.jctc.3c01421, D4SC06640E, Zheng_2026} have enabled \emph{hybrid} ML/MM force fields that predict bonded parameters ($\Phi_{\text{bonded}}$) directly from reference energies and forces, or model environmental polarization~\cite{song2025nepoip}. However, these models still rely on external, empirical nonbonded parameters ($\Phi_{\text{nonbonded}}$) or separate machine-learned and classical descriptions of the simulated system, limiting their ability to generalize across configurations. For instance, partial charges are typically derived from computationally expensive semi-empirical methods such as AM1-BCC~\citep{am1bcc2002} or by partitioning the electron density using methods such as MBIS~\citep{mbis2016}. Lennard-Jones (vdW) parameters, on the other hand, are assigned using rigid, curated typing rules inherited from legacy force fields including GAFF~\citep{gaff2004} and OpenFF~\cite{openff_wang2024}. 

In the following, we introduce  \texttt{grappa-fullFF}, a unified end-to-end approach that learns both bonded and nonbonded MM parameters from ab initio reference data. We extend Grappa~\cite{grappa2025} (hereafter referred to as  \texttt{grappa-bonded}), which has shown state-of-the-art performance on bonded parameters, to jointly predict all parameters. Instead of targeting each parameter independently, we train the model directly against QM energies and forces, allowing all components of the force field to be learned consistently. We regularize the implicitly learned MM parameters using physically motivated constraints on the electrostatic potential and the Lennard-Jones (LJ) parameters through supervision of the effective per-atom volume. 
We demonstrate that, even without explicit supervision of individual MM parameters, \gFF recovers the QM electrostatic potential with an accuracy comparable to expensive electron density partitioning schemes, without any electron density calculation at inference, and retains the performance of \gBOND  on a standard geometry optimization benchmark. 
In an explicit-solvent molecular dynamics (MD) simulation of alanine-dipeptide, our model reproduces the Ramachandran free-energy surface with accuracy comparable to \gBOND relative to AMBER99SB-ILDN \citep{10.1021/acs.jcim.5c01063}, and about twice as accurately as \texttt{espaloma-0.3.2}, even without having been trained on condensed phase systems.
\begin{figure}
    \centering
    \includegraphics[width=\linewidth]{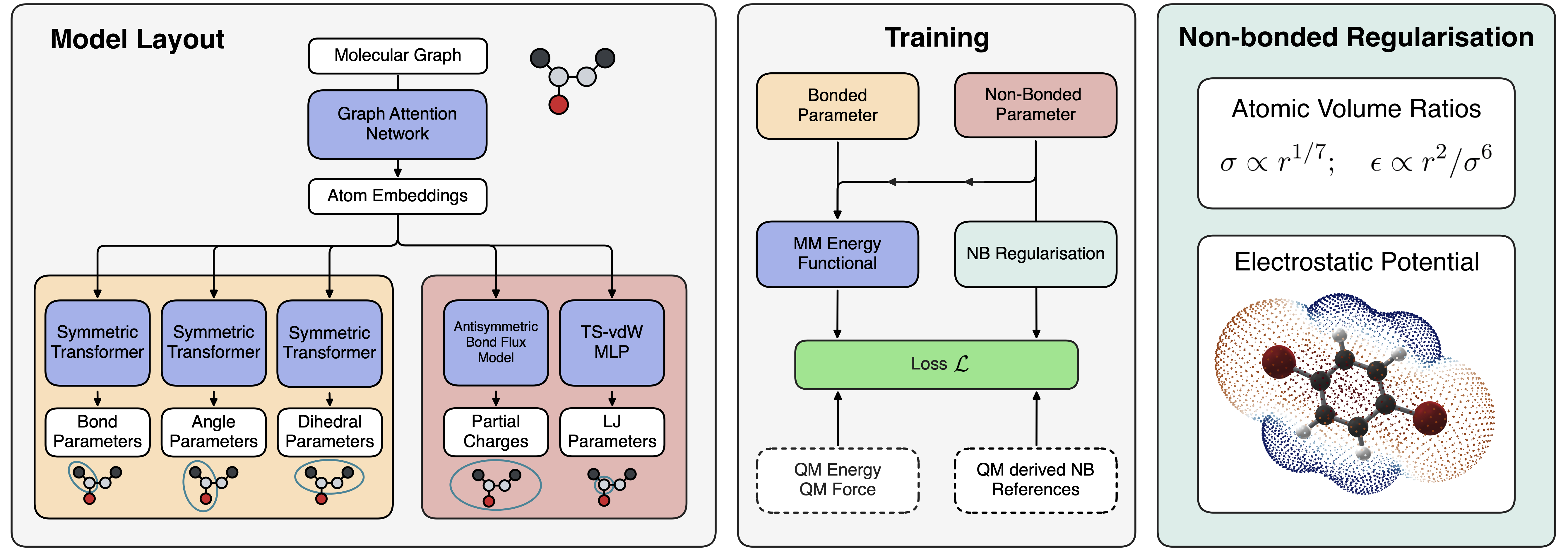}
    \caption{Grappa architecture and training routine.
    The shared graph-attention encoder feeds the atom embeddings in the individual parameter prediction heads.
    All parameters are trained simultaneously on QM energies and forces.
    The nonbonded parameter heads are additionally regularized by the electrostatic potential (ESP) and the ratio of the effective and free atomic volume, $v_i$.
    }
    \label{fig:grappa-model-workflow}
    \vspace{-1.5em}
\end{figure}

\section{Methodology}
Training the model by constraining energies and forces does not uniquely determine how the potential is decomposed into individual components of the FF in Eq.~\eqref{eq:mm_functional}. Various combinations of bonded and nonbonded interactions can yield similar total energies and forces, and similarly, different combinations of the electrostatic and vdW parameters can yield the same nonbonded contribution. This ambiguity, referred to as \emph{energy degeneracy}~\cite{wang2025design, 10.1021/acs.chemrev.4c00893}, can allow the model to achieve low errors on the target energies while implicitly learning parameters that are not physically meaningful and do not transfer reliably across molecular environments~\citep{Unke21}. 
We address this ambiguity by incorporating additional physical information, discussed in Sec.~\ref{sec:training-and-regularize}, which directly constrains the nonbonded interactions. Before introducing these constraints, we first describe the model architecture used to predict the MM parameters.

\subsection{Model Architecture}
We extend the \gBOND framework \citep{grappa2025}, which uses a graph attention network to predict the bonded parameters (Fig. \ref{fig:grappa-model-workflow}, shaded yellow), by introducing two new prediction heads for nonbonded interactions (shaded red), namely the antisymmetric bond-charge-flux head for partial charges and a volume-ratio head for LJ parameters. The parameterization of the bonded interactions is unchanged.
\paragraph{Antisymmetric Bond-Charge-Flux Head.}
To get the electrostatic contribution to the nonbonded energy, the model must learn to assign a partial charge to each atom. Each atomic charge state is initialized ($t=0$) as $q_{i}^{(t=0)}=Q_{\rm tot}/N$ and is iteratively refined through $t=1,\ldots,T$ message-passing steps. Instead of predicting atomic charges at each step, the model outputs charge flux between bonded atom pairs, $w_{ij}^{(t)}$, which denotes the charge transferred from atom $j$ to atom $i$. The fluxes are restricted to be antisymmetric under the permutation of atom indices, $w_{ij}^{(t)} = -w_{ji}^{(t)},$ so that the total charge $Q_{\rm tot}$ is conserved. 
Following the classical Bond-Charge Increment (BCI)~\cite{MMFF94}, we compute the atomic charges using the flux predictions from all bonded neighbors of $i$ (denoted as $\mathcal{N}(i)$) as, 
\begin{equation}
\label{eq:bci}
q_i^{(t+1)} = q_i^{(t)} + \sum_{j \in N(i)}  \left (w_{ij}^{(t)} - w_{ji}^{(t)}\right).
\end{equation}
The initial charge fluxes $w_{ij}^{(t=0)}$ are computed using the latent atom embeddings $h_i, h_j$, 
\begin{equation}
    w_{ij}^{(0)} = \frac{1}{2}\bigl[\mathrm{MLP}^{(0)}(h_i, h_j) - \mathrm{MLP}^{(0)}(h_j, h_i)\bigr]
\end{equation}
whereas, subsequent time steps additionally condition the fluxes on the current atomic charge state and its local change $\Delta q_{i,t} = q_{i,t} - q_{i,t-1}$. Denoting these inputs as $\bxi_{ij}^{(t)} = [h_i, h_j, q_{i}^{(t)}, q_{j}^{(t)}, \Delta q_{i}^{(t)}, \Delta q_{j}^{(t)}]$, nonlinear activation as $\sigma$, and element-wise product as $\odot$, the update for $t\geq 1$ can be written as,
\begin{equation}
    w_{ij}^{(t)} = \sigma\bigl(\mathrm{MLP}_{\text{gate}}^{(t)}(h_i + h_j, h_i \odot h_j\bigr) \cdot \frac{1}{2}\bigl[\mathrm{MLP}^{(t)}(\bxi_{ij}^{(t)}) - \mathrm{MLP}^{(t)}(\bxi_{ji}^{(t)})\bigr].
\end{equation}
The anti-symmetry of the charge fluxes enforces $\sum_i \Delta q_i^{(t)} = 0$ at every refinement step and maintains total charge conservation by construction, eliminating the need for post-hoc normalization or alternative charge equilibration~\citep{Rappe91, Gilson03}.

\paragraph{Volume-ratio LJ Head.}
Similarly, instead of directly predicting the LJ parameters $(\sigma, \epsilon)$, a separate MLP predicts per-atom volume ratios $v_i = v_i^\text{eff}/v_i^\text{free}$ from the latent atomic representation $h_i$.
This formulation couples $\sigma$ and $\epsilon$ through a physically motivated relationship, constraining their predicted values to be more meaningful. 
Following the extension~\citep{PhysRevLett.121.183401} of the TS-vdW formalism \citep{PhysRevLett.102.073005}, which uses $v_i$ to determine effective-atom polarizabilities $\alpha_\text{eff}= v_i \, \alpha_\text{free}$ and dispersion coefficients $C_6^\text{eff} = v_i^2 \,C_{6}^\text{free}$, we map LJ parameters to $v_i$ through $\sigma \propto \alpha_\text{eff}^{1/7}$ and $\epsilon \propto C_6^\text{eff}/\sigma^6$. 
For physical validity, we restrict the predicted volume ratio to be strictly positive using a scaled Exponential Linear Unit (ELU) activation on the network output.  

\subsection{Model training through differentiable MM and physics-informed regularization}
\label{sec:training-and-regularize}
To train the model on energies and forces, we pass the predicted bonded parameters, charges, and LJ parameters to a fully differentiable implementation of the energy functional (Eq. \eqref{eq:mm_functional}), thus allowing loss gradients to propagate through the parameter prediction and graph embeddings. The corresponding atomic forces are obtained as $\mathbf{F}=-\nabla_{\mathbf{x}}E_\text{MM}$. Our differentiable functional explicitly implements classical 1-2 and 1-3 exclusion rules, as well as 1-4 scaling factors, 
to mimic the deployment environment of standard MD engines (e.g., GROMACS\citep{gromacs2015}, OpenMM \citep{openmm8_2023}). 

As previously described, optimizing only the QM energy and force errors leaves the energy decomposition underconstrained. We therefore introduce additional physical constraints to the training objective and define the loss function as,  
\begin{align}
    \mathcal{L} = \lambda_E \mathcal{L}_E + \lambda_\mathbf{F} \mathcal{L}_\mathbf{F} + \lambda_\text{ESP} \mathcal{L}_\text{ESP} + \lambda_{v} \mathcal{L}_{v},
    \label{eq:loss}
\end{align}
where $\mathcal{L}_E$ and $\mathcal{L}_\mathbf{F}$ are the mean squared errors (MSEs) of the total MM energy and atomic forces relative to the QM references. $\mathcal{L}_\text{ESP}$ and $\mathcal{L}_{v}$ regularize the predicted charges and LJ parameters, described below, and $\lambda$s determine the relative weight of each term. 
Instead of optimizing the partial charges against their target values in eq.\eqref{eq:loss}, which would depend on the choice of partitioning scheme, we constrain them by matching their electrostatic potential (ESP) to the QM reference. We compare this choice against direct charge supervision in Appendix \ref{sec:app.ablation}. We evaluate the ESP at grid points $\boldsymbol{\chi}$ (which depend on $\bx$ implicitly) surrounding the molecule and define 
\begin{align}
    \mathcal{L}_\text{ESP} =\frac{1}{N_{\boldsymbol{\chi}}} \sum_{\text{shell}} \sum_{j \in \text{shell}} \left( \frac{\tilde{V}_\text{ESP}({ \{\tilde{q}_i\}},\boldsymbol{\chi}_{j})-V_\text{ESP}(\boldsymbol{\chi}_j)}{\mu_{\text{shell}}}
\right)^2,
\label{eq:esp_loss}
\end{align}
where $\tilde{V}$, $V$ are the predicted and reference ESP and  $\{\tilde{q}_i\}$ is the set of predicted atomic charges. Note that unlike $\tilde{V}$, the reference ESP does not rely on charges and is computed directly from the electron density. Grid points $\boldsymbol{\chi}_j$ are obtained using concentric Fibonacci lattice shells at 1.4, 1.6, 1.8, and 2 times the vdW radii around each atom, following conventional ESP fitting schemes\citep{Besler1990, Bayly93}, and grid points that fall inside the vdW radii of another atom are excluded. Contributions from each grid point are normalized by $\mu_{\text{shell}}$, defined as the root-mean-square of the reference ESP over the corresponding shell, to compensate for the rapid decay of the ESP with distance. 

For the LJ parameters, we similarly constrain the predicted effective volume ratios $\tilde{v}_i$ using reference ratios $v_i$ obtained by partitioning the QM electron density, and define
$\mathcal{L}_{v} = \frac{1}{N_{\rm atoms}} \sum_{i \in \bx} \left(\tilde{v}_i-v_i\right)^2.$
We use the losses on energies and forces to train all model parameters but restrict $\mathcal{L}_\text{ESP}$ and $\mathcal{L}_{v}$ to exclusively update the parameters of the charge flux and volume ratio prediction heads. 

\section{Results}
\label{sec:results}
In the following, we consider the DES-monomer, dipeptide, and PubChem subsets of SPICE \citep{eastman2022spicedatasetdruglikemolecules} using the QM energy and forces reported in OMol25 at the $\omega$B97M-V/def2-TZVPD level 
\citep{levine2026openmolecules2025omol25}. For direct comparison with Espaloma \citep{espaloma2022} and the \gBOND \citep{grappa2025}, we similarly filter high-force conformations, retaining $10{,}240$ unique molecules and ${\sim}441{,}000$ conformers for which QM energy and force labels are available.
Of these, electron densities are available for only $14{,}600$ conformers in the OMol25 dataset. Therefore, while all conformers contribute to the energy and force losses ($\mathcal{L}_E$, $\mathcal{L}_\mathbf{F}$), $\mathcal{L}_\text{ESP}$ and $\mathcal{L}_{v}$ are evaluated only for a subset of the dataset with electron densities.
Reference ESP values are computed using \texttt{Multiwfn} \citep{multiwfn2012, multiwfn2024}, while effective and free atomic volumes are obtained using MBIS partitioning \citep{mbis2016} of the available electron densities and averaged across all conformers per molecule to a single value per atom.

\paragraph{Accuracy of the Learned Electrostatics} We assess the quality of the learned charges by evaluating how well they reproduce the QM electric response properties. Table~\ref{tab:force_field_esp_dipole_quad_comparison} compares the RMSE of the ESP and dipole moments across the test dataset partition. \gFF outperforms traditional empirical charge schemes such as AM1-BCC and stand-alone machine-learned charge models such as \texttt{espaloma-charge} \citep{eastman2022spicedatasetdruglikemolecules}. Fig.~\ref{fig:esp_comp} shows deviations in the ESP computed from the predicted charges by different charge schemes from the QM reference. Notably, the predicted charges from \gFF are almost as accurate as those obtained from MBIS partitioning, even though the model is never trained against reference atomic charges and only infers them from ESP supervision.
\begin{table}[!h]
\centering
\caption{Comparison of ESP ($10^{-3} e/\text{\AA}$) and dipole moment ($e\text{\AA}$) RMSE across charge models and datasets. Parenthetical values give the standard deviation on the last digits.
}
\label{tab:force_field_esp_dipole_quad_comparison}
\resizebox{\textwidth}{!}{%
\begin{tabular}{l
    S[table-format=2.3(2)] S[table-format=1.3(2)]
    S[table-format=2.2(2)] S[table-format=1.3(2)]
    S[table-format=2.2(2)] S[table-format=1.3(2)]
}
\toprule
 & \multicolumn{2}{c}{PubChem} & \multicolumn{2}{c}{Dipeptide} & \multicolumn{2}{c}{DES Monomers} \\
\cmidrule(lr){2-3} \cmidrule(lr){4-5} \cmidrule(lr){6-7}
{Charge Model} & {ESP} & {Dipole} & {ESP} & {Dipole} & {ESP} & {Dipole}\\
\midrule
MBIS\footnotemark[1]         & \textbf{7.157(54)}  & 1.190(26) & \textbf{8.47(37)}  & 1.724(79) & 6.22(25) & 0.622(50)\\
AM1-BCC\footnotemark[1]          & 10.414(87) & 1.156(25) & 12.32(54) & 1.88(10) & 8.50(40) & 0.597(47) \\
\texttt{espaloma-charge}  & 35.7(10)   & 2.084(40) & 30.3(11)  & 2.174(71) & 18.2(17) & 0.758(66) \\
\gFF     & 8.90(58)   & \textbf{1.048(24)} & 12.7(13)  & \textbf{1.584(81)} & \textbf{6.12(26)} & \textbf{0.561(46)}\\
\bottomrule
\end{tabular}}
\vspace{-1em}
\end{table}
\footnotetext[1]{{Require DFT calculations}}
\paragraph{Structural Accuracy in Geometry Optimization} We next assess whether the learned force field reproduces QM equilibrium structures using a standard geometry optimization benchmark~\citep{openff_benchmark2020,takaba2023machinelearned}. Starting from unrelaxed conformations, we optimize each molecule and compare the resulting geometries and relative energies with the QM reference in Table~\ref{tab:force_field_comparison}. The distributions of geometry RMSE, TFD, and $|\Delta\Delta E|$ (ddE) in Fig.~\ref{fig:geom_benchmark} show a clear separation between ML and classical force fields. \texttt{espaloma-0.3.2},  \texttt{grappa-bonded}, and \gFF consistently outperform GAFF-2.11 and OpenFF-2.1.0 across all three metrics. Among ML models, the distributions are nearly indistinguishable, with marginal differences in the median values. This shows that \gFF retains the geometric fidelity of its bonded-only counterpart, and that jointly learning nonbonded parameters does not degrade the bonded parameters due to energy degeneracy.
\begin{table}[!h]
\centering
\caption{Comparison of force field performance metrics: median of the Root Mean Square Error (RMSE med) in geometries between MM-optimized and QM-optimized conformers, Torsion Fingerprint Deviation (TFD), and the Mean Absolute Error (MAE) and RMSE of the conformer-averaged relative energy (ddE).
}
\label{tab:force_field_comparison}
\begin{tabularx}{\textwidth}{l *{4}{>{\raggedleft\arraybackslash}X}}
\toprule
Force Field & RMSE med (\AA) & TFD med & ddE MAE & ddE RMSE \\
\midrule
gaff-2.11              & 0.4997 & 0.0476 & 2.341 & 3.892 \\
openff-2.1.0            & 0.3688 & 0.0355 & 2.036 & 3.495 \\
\texttt{espaloma-0.3.2}          & 0.2725 & 0.0241 & 1.639 & 3.064 \\
\gBOND   & 0.2571 & 0.0232 & 1.608 & 2.773 \\
\gFF  & 0.2661 & 0.0221 & 1.589 & 2.973 \\
\bottomrule
\end{tabularx}
\vspace{-1em}
\end{table}

\paragraph{Conformational Thermodynamics in Explicit Solvent}\label{sec:res.md} Finally, we evaluate \gFF in explicit-solvent molecular dynamics of capped alanine dipeptide (ACE-ALA-NME). We run $500$~ns of unbiased MD with TIP3P water using AMBER99SB-ILDN, \texttt{espaloma-0.3.2},  \texttt{grappa-bonded}, and  \texttt{grappa-fullFF}, and compare the resulting Ramachandran free-energy surfaces (FES). While \gBOND and \texttt{espaloma-0.3.2} combine their learned bonded parameters with empirical or separately learned nonbonded parameters, \gFF predicts both bonded and nonbonded parameters within a single model. Full simulation details are given in Appendix~\ref{sec:app.md-protocol}. 
All four force fields recover the characteristic extended/PPII/$\alpha_R$ basin as the dominant low-free-energy region and $\alpha_L$ as a distinct, higher-energy minimum. To quantify the differences, we bin the two backbone dihedral angles $(\phi,\psi)$ on a $5^\circ$ grid and compute the symmetric Kullback--Leibler (KL) and Jensen--Shannon (JS) divergences relative to AMBER99SB-ILDN (Table~\ref{tab:rama_kl}). Both Grappa models achieve $\mathrm{JS}\lesssim0.08$~nats, approximately twice as close to AMBER99SB-ILDN as \texttt{espaloma-0.3.2}, and agree with each other, as the difference in their JS is only $\Delta\mathrm{JS}=0.013$~nats. Thus, our unified training preserves the conformational sampling of  \texttt{grappa-bonded}, and eliminates the need to assign nonbonded parameters separately. 
\begin{figure}[!htb]
    \centering
    \vspace{-0.5em}
    \includegraphics[width=\linewidth]{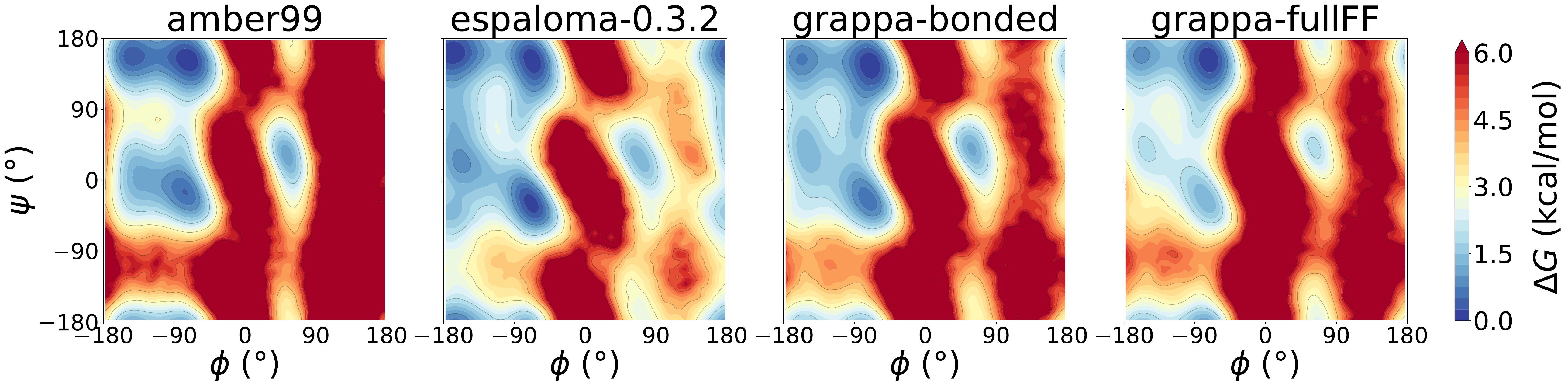}
    \vspace{-1.5em}
    \caption{Ramachandran FES of capped Alanine dipeptide from $500$ ns explicit-solvent (TIP3P) MD.}
    \label{fig:md-fes}
    \vspace{-1em}
\end{figure}

\section{Discussion and Conclusions}
Our results demonstrate that \gFF is an end-to-end force field that learns bonded and nonbonded parameters directly from ab-initio QM energies and forces. 
One might expect this approach to introduce a degeneracy between interaction terms, but we show that this ambiguity can be resolved through physics-inspired regularization. By constraining charge-conserving bond fluxes through the electrostatic potential and LJ parameters through the TS-vdW volume ratios, our model learns parameters which are physically consistent and transferable. For instance, we obtain reduced LJ $\sigma$ values for polar hydrogens required for hydrogen bonding, despite the larger values that would result from using TS-vdW volume ratios directly (Appendix ~\ref{sec:app.H-TS-vdW}). 
These nonbonded parameters (charges, LJ parameters) emerge implicitly, without having been optimized against reference values, and offer considerably cheaper alternatives to methods such as MBIS that rely on expensive electron density calculations and partitioning. 
This parameter flexibility does not compromise the accuracy of bonded interactions. \gFF matches the state-of-the-art performance of \gBOND on geometry optimization and, despite being trained exclusively on isolated gas-phase monomers, reproduces conformational thermodynamics in explicit solvent. 
Extending training to explicit-solvent configurations will be a next step toward further refining model accuracy, along with additional refinement steps or alternative charge initialization schemes for larger, more delocalized molecules, paving the way for high-throughput, closed-loop simulations.

\begin{ack}
This project has received funding from the European Research Council (ERC) under the European Union’s Horizon 2020 research and innovation programme (grant agreement No. 101002812)

Computations were performed on the HPC system Otter at the Max Planck Computing and Data Facility.

The Flatiron Institute is a
division of the Simons Foundation. 


\end{ack}

\bibliographystyle{unsrtnat}
\bibliography{bib.bib}

@article{grappa2025,
    author = {Seute, Leif and Hartmann, Eric and Stühmer, Jan and Gräter, Frauke},
    title = {Grappa – a machine learned molecular mechanics force field},
    journal = {Chemical Science},
    volume = {16},
    number = {6},
    pages = {2907-2930},
    year = {2025},
    month = {02},
    issn = {2041-6520},
    doi = {10.1039/d4sc05465b},
    url = {https://doi.org/10.1039/d4sc05465b},
}

@misc{eastman2022spicedatasetdruglikemolecules,
      title={SPICE, A Dataset of Drug-like Molecules and Peptides for Training Machine Learning Potentials}, 
      author={Peter Eastman and Pavan Kumar Behara and David L. Dotson and Raimondas Galvelis and John E. Herr and Josh T. Horton and Yuezhi Mao and John D. Chodera and Benjamin P. Pritchard and Yuanqing Wang and Gianni De Fabritiis and Thomas E. Markland},
      year={2022},
      eprint={2209.10702},
      archivePrefix={arXiv},
      primaryClass={physics.chem-ph},
      url={https://arxiv.org/abs/2209.10702}, 
}

@misc{levine2026openmolecules2025omol25,
      title={The Open Molecules 2025 (OMol25) Dataset, Evaluations, and Models}, 
      author={Daniel S. Levine and Muhammed Shuaibi and Evan Walter Clark Spotte-Smith and Michael G. Taylor and Muhammad R. Hasyim and Kyle Michel and Ilyes Batatia and Gábor Csányi and Misko Dzamba and Peter Eastman and Nathan C. Frey and Xiang Fu and Vahe Gharakhanyan and Aditi S. Krishnapriyan and Joshua A. Rackers and Sanjeev Raja and Ammar Rizvi and Andrew S. Rosen and Zachary Ulissi and Santiago Vargas and C. Lawrence Zitnick and Samuel M. Blau and Brandon M. Wood},
      year={2026},
      eprint={2505.08762},
      archivePrefix={arXiv},
      primaryClass={physics.chem-ph},
      url={https://arxiv.org/abs/2505.08762}, 
}

@article{Bayly93,
author = {Bayly, Christopher I. and Cieplak, Piotr and Cornell, Wendy and Kollman, Peter A.},
title = {A well-behaved electrostatic potential based method using charge restraints for deriving atomic charges: the RESP model},
journal = {The Journal of Physical Chemistry},
volume = {97},
number = {40},
pages = {10269-10280},
year = {1993},
doi = {10.1021/j100142a004},
URL ={https://doi.org/10.1021/j100142a004}
}

@article{Unke21,
author = {Unke, Oliver T. and Chmiela, Stefan and Sauceda, Huziel E. and Gastegger, Michael and Poltavsky, Igor and Sch{\"u}tt, Kristof T. and Tkatchenko, Alexandre and M{\"u}ller, Klaus-Robert},
title = {Machine Learning Force Fields},
journal = {Chemical Reviews},
volume = {121},
number = {16},
pages = {10142-10186},
year = {2021},
doi = {10.1021/acs.chemrev.0c01111},
    note ={PMID: 33705118},
}

@article{Rappe91,
author = {Rappe, Anthony K. and Goddard, William A. III},
title = {Charge equilibration for molecular dynamics simulations},
journal = {The Journal of Physical Chemistry},
volume = {95},
number = {8},
pages = {3358-3363},
year = {1991},
doi = {10.1021/j100161a070},
}

@article{Gilson03,
author = {Gilson, Michael K. and Gilson, Hillary S. R. and Potter, Michael J.},
title = {Fast Assignment of Accurate Partial Atomic Charges: An Electronegativity Equalization Method that Accounts for Alternate Resonance Forms},
journal = {Journal of Chemical Information and Computer Sciences},
volume = {43},
number = {6},
pages = {1982-1997},
year = {2003},
doi = {10.1021/ci034148o},
    note ={PMID: 14632449},
}

@article{PhysRevLett.102.073005,
  title = {Accurate Molecular Van Der Waals Interactions from Ground-State Electron Density and Free-Atom Reference Data},
  author = {Tkatchenko, Alexandre and Scheffler, Matthias},
  journal = {Phys. Rev. Lett.},
  volume = {102},
  issue = {7},
  pages = {073005},
  numpages = {4},
  year = {2009},
  month = {Feb},
  publisher = {American Physical Society},
  doi = {10.1103/PhysRevLett.102.073005},
  url = {https://link.aps.org/doi/10.1103/PhysRevLett.102.073005}
}

@article{MMFF94,
author = {Halgren, Thomas A.},
title = {Merck molecular force field. I. Basis, form, scope, parameterization, and performance of MMFF94},
journal = {Journal of Computational Chemistry},
volume = {17},
number = {5-6},
pages = {490-519},
doi = {10.1002/(SICI)1096-987X(199604)17:5/6<490::AID-JCC1>3.0.CO;2-P},
url = {https://onlinelibrary.wiley.com/doi/abs/10.1002/%28SICI%291096-987X%28199604%2917%3A5/6%3C490%3A%3AAID-JCC1%3E3.0.CO%3B2-P},
year = {1996}
}

@article{wang2025design,
  title={On the design space between molecular mechanics and machine learning force fields},
  author={Wang, Yuanqing and Takaba, Kenichiro and Chen, {Michael S} and Wieder, Marcus and Xu, Yuzhi and Zhu, Tong and Zhang, {John ZH} and Nagle, Arnav and Yu, Kuang and Wang, Xinyan and others},
  journal={Applied Physics Reviews},
  volume={12},
  number={2},
  year={2025},
  publisher={AIP Publishing}
}

@article{espaloma2022,
   title={End-to-end differentiable construction of molecular mechanics force fields},
   volume={13},
   ISSN={2041-6539},
   url={http://dx.doi.org/10.1039/D2SC02739A},
   DOI={10.1039/d2sc02739a},
   number={41},
   journal={Chemical Science},
   publisher={Royal Society of Chemistry (RSC)},
   author={Wang, Yuanqing and Fass, Josh and Kaminow, Benjamin and Herr, John E. and Rufa, Dominic and Zhang, Ivy and Pulido, Iván and Henry, Mike and Bruce Macdonald, Hannah E. and Takaba, Kenichiro and Chodera, John D.},
   year={2022},
   pages={12016–12033} }

@article{am1bcc2002,
author = {Jakalian, Araz and Jack, David B. and Bayly, Christopher I.},
title = {Fast, efficient generation of high-quality atomic charges. AM1-BCC model: II. Parameterization and validation},
journal = {Journal of Computational Chemistry},
volume = {23},
number = {16},
pages = {1623-1641},
doi = {https://doi.org/10.1002/jcc.10128},
url = {https://onlinelibrary.wiley.com/doi/abs/10.1002/jcc.10128},
eprint = {https://onlinelibrary.wiley.com/doi/pdf/10.1002/jcc.10128},
year = {2002}
}

@article{gaff2004,
author = {Wang, Junmei and Wolf, Romain M. and Caldwell, James W. and Kollman, Peter A. and Case, David A.},
title = {Development and testing of a general amber force field},
journal = {Journal of Computational Chemistry},
volume = {25},
number = {9},
pages = {1157-1174},
doi = {https://doi.org/10.1002/jcc.20035},
url = {https://onlinelibrary.wiley.com/doi/abs/10.1002/jcc.20035},
eprint = {https://onlinelibrary.wiley.com/doi/pdf/10.1002/jcc.20035},
year = {2004}
}

@article{openff_benchmark2020,
author = {Lim, Victoria and Hahn, David and Tresadern, Gary and Bayly, Christopher and Mobley, David},
year = {2020},
month = {12},
pages = {1390},
title = {Benchmark assessment of molecular geometries and energies from small molecule force fields},
volume = {9},
journal = {F1000Research},
doi = {10.12688/f1000research.27141.1}
}

@misc{takaba2023machinelearned,
      title={Machine-learned molecular mechanics force field for the simulation of protein-ligand systems and beyond}, 
      author={Takaba, Kenichiro and Pulido, Iv{\`a}n and Behara, Pavan Kumar and Cavender, Chapin E. and Friedman, Anika J. and Henry, Michael M. and Opeskin, Hugo MacDermott and Iacovella, Christopher R. and Nagle, Arnav M. and Payne, Alexander Matthew and Shirts, Michael R. and Mobley, David L. and Chodera, John D. and Wang, Yuanqing},
      year={2023},
      eprint={2307.07085},
      archivePrefix={arXiv},
      primaryClass={physics.chem-ph}
}

@article{10.1039/d4sc00690a,
    author = {Takaba, Kenichiro and Friedman, Anika J. and Cavender, Chapin E. and Behara, Pavan Kumar and Pulido, Iván and Henry, Michael M. and MacDermott-Opeskin, Hugo and Iacovella, Christopher R. and Nagle, Arnav M. and Payne, Alexander Matthew and Shirts, Michael R. and Mobley, David L. and Chodera, John D. and Wang, Yuanqing},
    title = {Machine-learned molecular mechanics force fields from large-scale quantum chemical data},
    journal = {Chemical Science},
    volume = {15},
    number = {32},
    pages = {12861-12878},
    year = {2024},
    month = {08},
    issn = {2041-6520},
    doi = {10.1039/d4sc00690a},
    url = {https://doi.org/10.1039/d4sc00690a},
    eprint = {https://pubs.rsc.org/sc/article-pdf/15/32/12861/10027570/d4sc00690a.pdf},
}

@article{mbis2016,
    author = {Verstraelen, Toon and Vandenbrande, Steven and Heidar-Zadeh, Farnaz and Vanduyfhuys, Louis and Van Speybroeck, Veronique and Waroquier, Michel and Ayers, Paul W.},
    title = {Minimal Basis Iterative Stockholder: Atoms in Molecules
for Force-Field Development},
    journal = {Journal of Chemical Theory and Computation},
    volume = {12},
    number = {8},
    pages = {3894-3912},
    year = {2016},
    month = {07},
    issn = {1549-9618},
    doi = {10.1021/acs.jctc.6b00456},
    url = {https://doi.org/10.1021/acs.jctc.6b00456},
    eprint = {https://pubs.acs.org/jctcce/article-pdf/12/8/3894/13467585/ct6b00456.pdf},
}

@article{10.1021/acs.chemrev.4c00893,
    author = {Tretiakov, Serhii and Nigam, AkshatKumar and Pollice, Robert},
    title = {Studying Noncovalent
Interactions in Molecular Systems
with Machine Learning},
    journal = {Chemical Reviews},
    volume = {125},
    number = {12},
    pages = {5776-5829},
    year = {2025},
    month = {06},
    issn = {0009-2665},
    doi = {10.1021/acs.chemrev.4c00893},
    url = {https://doi.org/10.1021/acs.chemrev.4c00893},
    eprint = {https://pubs.acs.org/chreay/article-pdf/125/12/5776/41212999/cr4c00893.pdf},
}

@article{multiwfn2012,
author = {Lu, Tian and Chen, Feiwu},
title = {Multiwfn: A multifunctional wavefunction analyzer},
journal = {Journal of Computational Chemistry},
volume = {33},
number = {5},
pages = {580-592},
doi = {https://doi.org/10.1002/jcc.22885},
url = {https://onlinelibrary.wiley.com/doi/abs/10.1002/jcc.22885},
eprint = {https://onlinelibrary.wiley.com/doi/pdf/10.1002/jcc.22885},
year = {2012}
}

@article{multiwfn2024,
    author = {Lu, Tian},
    title = {A comprehensive electron wavefunction analysis toolbox for chemists, Multiwfn},
    journal = {The Journal of Chemical Physics},
    volume = {161},
    number = {8},
    pages = {082503},
    year = {2024},
    month = {08},
    issn = {0021-9606},
    doi = {10.1063/5.0216272},
    url = {https://doi.org/10.1063/5.0216272},
    eprint = {https://pubs.aip.org/aip/jcp/article-pdf/doi/10.1063/5.0216272/20129571/082503_1_5.0216272.pdf},
}

@article{Besler1990,
author = {Besler, Brent H. and Merz Jr., Kenneth M. and Kollman, Peter A.},
title = {Atomic charges derived from semiempirical methods},
journal = {Journal of Computational Chemistry},
volume = {11},
number = {4},
pages = {431-439},
doi = {https://doi.org/10.1002/jcc.540110404},
url = {https://onlinelibrary.wiley.com/doi/abs/10.1002/jcc.540110404},
eprint = {https://onlinelibrary.wiley.com/doi/pdf/10.1002/jcc.540110404},
year = {1990}
}

@article{openff_wang2024,
    author = {Wang, Lily and Behara, Pavan Kumar and Thompson, Matthew W. and Gokey, Trevor and Wang, Yuanqing and Wagner, Jeffrey R. and Cole, Daniel J. and Gilson, Michael K. and Shirts, Michael R. and Mobley, David L.},
    title = {The Open Force
Field Initiative: Open Software and
Open Science for Molecular Modeling},
    journal = {The Journal of Physical Chemistry B},
    volume = {128},
    number = {29},
    pages = {7043-7067},
    year = {2024},
    month = {07},
    issn = {1520-6106},
    doi = {10.1021/acs.jpcb.4c01558},
    url = {https://doi.org/10.1021/acs.jpcb.4c01558},
    eprint = {https://pubs.acs.org/jpcbfk/article-pdf/128/29/7043/12117346/jp4c01558.pdf},
}

@article{PhysRevLett.121.183401,
  title = {Quantum-Mechanical Relation between Atomic Dipole Polarizability and the van der Waals Radius},
  author = {Fedorov, Dmitry V. and Sadhukhan, Mainak and St\"ohr, Martin and Tkatchenko, Alexandre},
  journal = {Phys. Rev. Lett.},
  volume = {121},
  issue = {18},
  pages = {183401},
  numpages = {6},
  year = {2018},
  month = {Nov},
  publisher = {American Physical Society},
  doi = {10.1103/PhysRevLett.121.183401},
  url = {https://link.aps.org/doi/10.1103/PhysRevLett.121.183401}
}

@article{shaw2010atomic,
  title={Atomic-level characterization of the structural dynamics of proteins},
  author={Shaw, David E and Maragakis, Paul and Lindorff-Larsen, Kresten and Piana, Stefano and Dror, Ron O and Eastwood, Michael P and Bank, Joseph A and Jumper, John M and Salmon, John K and Shan, Yibing and others},
  journal={Science},
  volume={330},
  number={6002},
  pages={341--346},
  year={2010},
  publisher={American Association for the Advancement of Science}
}

@article{borhani2012future,
  title={The future of molecular dynamics simulations in drug discovery},
  author={Borhani, David W and Shaw, David E},
  journal={Journal of computer-aided molecular design},
  volume={26},
  number={1},
  pages={15--26},
  year={2012},
  publisher={Springer}
}

@article{morawietz2021machine,
  title={Machine learning-accelerated quantum mechanics-based atomistic simulations for industrial applications},
  author={Morawietz, Tobias and Artrith, Nongnuch},
  journal={Journal of Computer-Aided Molecular Design},
  volume={35},
  number={4},
  pages={557--586},
  year={2021},
  publisher={Springer}
}

@article{shibuta2017heterogeneity,
  title={Heterogeneity in homogeneous nucleation from billion-atom molecular dynamics simulation of solidification of pure metal: Molecular dynamics simulation of pure metal},
  author={Shibuta, Yasushi and Sakane, Shinji and Miyoshi, Eisuke and Okita, Shin and Takaki, Tomohiro and Ohno, Munekazu},
  journal={Nature communications},
  volume={8},
  number={1},
  pages={10},
  year={2017},
  publisher={Nature Publishing Group UK London}
}

@incollection{noe2020machine,
  title={Machine learning for molecular dynamics on long timescales},
  author={No{\'e}, Frank},
  booktitle={Machine learning meets quantum physics},
  pages={331--372},
  year={2020},
  publisher={Springer}
}

@article{10.1021/acs.jcim.5c01063,
    author = {Case, David
A. and Cerutti, David S. and Cruzeiro, Vinícius Wilian D. and Darden, Thomas A. and Duke, Robert E. and Ghazimirsaeed, Mahdieh and Giambaşu, George M. and Giese, Timothy J. and Götz, Andreas W. and Harris, Julie A. and Kasavajhala, Koushik and Lee, Tai-Sung and Li, Zhen and Lin, Charles and Liu, Jian and Miao, Yinglong and Salomon-Ferrrer, Romelia and Shen, Jana and Snyder, Ryan and Swails, Jason and Walker, Ross C. and Wang, Jinan and Wu, Xiongwu and Zeng, Jinzhe and Cheatham III, Thomas E. and Roe, Daniel R. and Roitberg, Adrian and Simmerling, Carlos and York, Darrin M. and Nagan, Maria C. and Merz, Kenneth M., Jr.},
    title = {Recent Developments
in Amber Biomolecular Simulations},
    journal = {Journal of Chemical Information and Modeling},
    volume = {65},
    number = {15},
    pages = {7835-7843},
    year = {2025},
    month = {07},
    issn = {1549-9596},
    doi = {10.1021/acs.jcim.5c01063},
    url = {https://doi.org/10.1021/acs.jcim.5c01063},
    eprint = {https://pubs.acs.org/jcisd8/article-pdf/65/15/7835/41162502/ci5c01063.pdf},
}

@article{10.1021/acs.jctc.3c01421,
    author = {Chen, Gong and Jaffrelot Inizan, Théo and Plé, Thomas and Lagardère, Louis and Piquemal, Jean-Philip and Maday, Yvon},
    title = {Advancing Force Fields Parameterization: A Directed
Graph Attention Networks Approach},
    journal = {Journal of Chemical Theory and Computation},
    volume = {20},
    number = {13},
    pages = {5558-5569},
    year = {2024},
    month = {06},
    issn = {1549-9618},
    doi = {10.1021/acs.jctc.3c01421},
    url = {https://doi.org/10.1021/acs.jctc.3c01421},
    eprint = {https://pubs.acs.org/jctcce/article-pdf/20/13/5558/9568534/ct3c01421.pdf},
}

@article{10.1021/ct9000685,
    author = {Harvey, M. J. and Giupponi, G. and Fabritiis, G. De},
    title = {ACEMD: Accelerating Biomolecular Dynamics in the Microsecond Time Scale},
    journal = {Journal of Chemical Theory and Computation},
    volume = {5},
    number = {6},
    pages = {1632-1639},
    year = {2009},
    month = {05},
    issn = {1549-9618},
    doi = {10.1021/ct9000685},
    url = {https://doi.org/10.1021/ct9000685},
    eprint = {https://pubs.acs.org/jctcce/article-pdf/5/6/1632/2209120/ct9000685.pdf},
}

@article{10.1371/journal.pcbi.1005659,
    doi = {10.1371/journal.pcbi.1005659},
    author = {Eastman, Peter AND Swails, Jason AND Chodera, John D. AND McGibbon, Robert T. AND Zhao, Yutong AND Beauchamp, Kyle A. AND Wang, Lee-Ping AND Simmonett, Andrew C. AND Harrigan, Matthew P. AND Stern, Chaya D. AND Wiewiora, Rafal P. AND Brooks, Bernard R. AND Pande, Vijay S.},
    journal = {PLOS Computational Biology},
    publisher = {Public Library of Science},
    title = {OpenMM 7: Rapid development of high performance algorithms for molecular dynamics},
    year = {2017},
    month = {07},
    volume = {13},
    url = {https://doi.org/10.1371/journal.pcbi.1005659},
    pages = {1-17},
    number = {7},

}

@article{10.1021/ct400314y,
    author = {Salomon-Ferrer, Romelia and Götz, Andreas W. and Poole, Duncan and Le Grand, Scott and Walker, Ross C.},
    title = {Routine Microsecond Molecular Dynamics Simulations
with AMBER on GPUs. 2. Explicit Solvent Particle Mesh Ewald},
    journal = {Journal of Chemical Theory and Computation},
    volume = {9},
    number = {9},
    pages = {3878-3888},
    year = {2013},
    month = {08},
    issn = {1549-9618},
    doi = {10.1021/ct400314y},
    url = {https://doi.org/10.1021/ct400314y},
    eprint = {https://pubs.acs.org/jctcce/article-pdf/9/9/3878/8192888/ct400314y.pdf},
}

@article{Hagler2019,
author = {Dauber-Osguthorpe, Pnina and Hagler, Arnie},
year = {2019},
month = {02},
pages = {},
title = {Biomolecular force fields: where have we been, where are we now, where do we need to go and how do we get there?},
volume = {33},
journal = {Journal of Computer-Aided Molecular Design},
doi = {10.1007/s10822-018-0111-4}
}

@article{DBLP:journals/jcamd/Hagler19,
  author       = {Arnold T. Hagler},
  title        = {Force field development phase {II:} Relaxation of physics-based criteria...
                  or inclusion of more rigorous physics into the representation of molecular
                  energetics},
  journal      = {J. Comput. Aided Mol. Des.},
  volume       = {33},
  number       = {2},
  pages        = {205--264},
  year         = {2019},
  url          = {https://doi.org/10.1007/s10822-018-0134-x},
  doi          = {10.1007/S10822-018-0134-X},
  bibsource    = {dblp computer science bibliography, https://dblp.org}
}

@article{723ed04a-9dcc-30ee-832b-e24d0998621e,
 ISSN = {09501207},
 URL = {http://www.jstor.org/stable/94265},
 author = {J. E. Jones},
 journal = {Proceedings of the Royal Society of London. Series A, Containing Papers of a Mathematical and Physical Character},
 number = {738},
 pages = {463--477},
 publisher = {The Royal Society},
 title = {On the Determination of Molecular Fields. II. From the Equation of State of a Gas},
 urldate = {2026-08-27},
 volume = {106},
 year = {1924}
}

@article{gromacs2015,
title = {GROMACS: High performance molecular simulations through multi-level parallelism from laptops to supercomputers},
journal = {SoftwareX},
volume = {1-2},
pages = {19-25},
year = {2015},
issn = {2352-7110},
doi = {https://doi.org/10.1016/j.softx.2015.06.001},
url = {https://www.sciencedirect.com/science/article/pii/S2352711015000059},
author = {Mark James Abraham and Teemu Murtola and Roland Schulz and Szilárd Páll and Jeremy C. Smith and Berk Hess and Erik Lindahl},
}

@article{openmm8_2023,
    author = {Eastman, Peter and Galvelis, Raimondas and Peláez, Raúl P. and Abreu, Charlles R. A. and Farr, Stephen E. and Gallicchio, Emilio and Gorenko, Anton and Henry, Michael M. and Hu, Frank and Huang, Jing and Krämer, Andreas and Michel, Julien and Mitchell, Joshua A. and Pande, Vijay S. and Rodrigues, João PGLM and Rodriguez-Guerra, Jaime and Simmonett, Andrew C. and Singh, Sukrit and Swails, Jason and Turner, Philip and Wang, Yuanqing and Zhang, Ivy and Chodera, John D. and De Fabritiis, Gianni and Markland, Thomas E.},
    title = {OpenMM 8: Molecular
Dynamics Simulation with Machine
Learning Potentials},
    journal = {The Journal of Physical Chemistry B},
    volume = {128},
    number = {1},
    pages = {109-116},
    year = {2023},
    month = {12},
    issn = {1520-6106},
    doi = {10.1021/acs.jpcb.3c06662},
    url = {https://doi.org/10.1021/acs.jpcb.3c06662},
    eprint = {https://pubs.acs.org/jpcbfk/article-pdf/128/1/109/15835521/jp3c06662.pdf},
}

@misc{batatia2023macehigherorderequivariant,
      title={MACE: Higher Order Equivariant Message Passing Neural Networks for Fast and Accurate Force Fields}, 
      author={Ilyes Batatia and Dávid Péter Kovács and Gregor N. C. Simm and Christoph Ortner and Gábor Csányi},
      year={2023},
      eprint={2206.07697},
      archivePrefix={arXiv},
      primaryClass={stat.ML},
      url={https://arxiv.org/abs/2206.07697}, 
}

@article{batzner2023nequip,
  title={E (3)-equivariant graph neural networks for data-efficient and accurate interatomic potentials},
  author={Batzner, Simon and Musaelian, Albert and Sun, Lixin and Geiger, Mario and Mailoa, Jonathan P and Kornbluth, Mordechai and Molinari, Nicola and Smidt, Tess E and Kozinsky, Boris},
  journal={Nature communications},
  volume={13},
  number={1},
  pages={2453},
  year={2022},
  publisher={Nature Publishing Group UK London}
}

@article{mazitov2025pet,
  title={PET-MAD as a lightweight universal interatomic potential for advanced materials modeling},
  author={Mazitov, Arslan and Bigi, Filippo and Kellner, Matthias and Pegolo, Paolo and Tisi, Davide and Fraux, Guillaume and Pozdnyakov, Sergey and Loche, Philip and Ceriotti, Michele},
  journal={Nature Communications},
  volume={16},
  number={1},
  pages={10653},
  year={2025},
  publisher={Nature Publishing Group UK London}
}

@article{bartok2010gaussian,
  title={Gaussian approximation potentials: The accuracy of quantum mechanics, without the electrons},
  author={Bart{\'o}k, Albert P and Payne, Mike C and Kondor, Risi and Cs{\'a}nyi, G{\'a}bor},
  journal={Physical review letters},
  volume={104},
  number={13},
  pages={136403},
  year={2010},
  publisher={APS}
}

@article{lysogorskiy2026graph,
  title={Graph atomic cluster expansion for foundational machine learning interatomic potentials},
  author={Lysogorskiy, Yury and Bochkarev, Anton and Drautz, Ralf},
  journal={npj Computational Materials},
  volume={12},
  number={1},
  pages={114},
  year={2026},
  publisher={Nature Publishing Group UK London}
}

@article{morado2023does,
  title={Does a machine-learned potential perform better than an optimally tuned traditional force field? a case study on fluorohydrins},
  author={Morado, Jo{\~a}o and Mortenson, Paul N and Nissink, J Willem M and Essex, Jonathan W and Skylaris, Chris-Kriton},
  journal={Journal of Chemical Information and Modeling},
  volume={63},
  number={9},
  pages={2810},
  year={2023}
}

@article{smith2017ani,
  title={ANI-1: an extensible neural network potential with DFT accuracy at force field computational cost},
  author={Smith, Justin S and Isayev, Olexandr and Roitberg, Adrian E},
  journal={Chemical science},
  volume={8},
  number={4},
  pages={3192--3203},
  year={2017},
  publisher={The Royal Society of Chemistry}
}

@article{song2025nepoip,
  title={NepoIP/MM: toward accurate biomolecular simulation with a machine learning/molecular mechanics model incorporating polarization effects},
  author={Song, Ge and Yang, Weitao},
  journal={Journal of Chemical Theory and Computation},
  volume={21},
  number={11},
  pages={5588--5598},
  year={2025},
  publisher={ACS Publications}
}

@Article{D4SC06640E,
  author    = {Tianze Zheng and Ailun Wang and Xu Han and Yu Xia and Xingyuan Xu and Jiawei Zhan and Yu Liu and Yang Chen and Zhi Wang and Xiaojie Wu and Sheng Gong and Wen Yan},
  title     = {Data-driven parametrization of molecular mechanics force fields for expansive chemical space coverage},
  journal   = {Chem. Sci.},
  year      = {2025},
  pages     = {-},
  publisher = {The Royal Society of Chemistry},
  doi       = {10.1039/D4SC06640E},
  url       = {http://dx.doi.org/10.1039/D4SC06640E}
}

@article{Zheng_2026,
   title={Bridging quantum mechanics to liquid properties via a universal organic force field},
   volume={17},
   ISSN={2041-1723},
   url={http://dx.doi.org/10.1038/s41467-026-73566-3},
   DOI={10.1038/s41467-026-73566-3},
   number={1},
   journal={Nature Communications},
   publisher={Springer Science and Business Media LLC},
   author={Zheng, Tianze and Xu, Xingyuan and Wang, Zhi and Yang, Zhenze and Wang, Yuanheng and Han, Xu and Chen, Lei and Mu, Zhenliang and Zhang, Ziqing and Liu, Siyuan and Gong, Sheng and Yu, Kuang and Yan, Wen},
   year={2026},
   month=May }

\newpage
\appendix

\section{Extended Benchmarking and Visualizations}
\subsection{Qualitative Analysis of Electrostatic Potential Surfaces}
Although Table \ref{tab:force_field_esp_dipole_quad_comparison} summarizes the ESP errors, visually inspecting the ESP projected onto the molecular van der Waals surface provides a more direct intuition for how these models will behave in intermolecular interactions. 
The 1.4$\times$ vdW surface is particularly relevant for nonbonded interactions, as it approximates the intermolecular contact distances. Figure \ref{fig:esp_comp} illustrates the ESP surface of a representative polar molecule.
\gFF closely reproduces both the spatial structure and magnitude of the QM reference, comparable to the more computationally expensive MBIS charges. In contrast, empirical or learned charges in AM1-BCC and \texttt{espaloma-0.3.2} show larger deviations in several regions of the molecular surface, often over- or under-polarizing with respect to QM.
These visualizations confirm that ESP supervision constrains the predicted partial charges to reproduce the electrostatics from the underlying ab initio electron density.
\begin{figure}[!h]
    \centering
    \includegraphics[width=\linewidth]{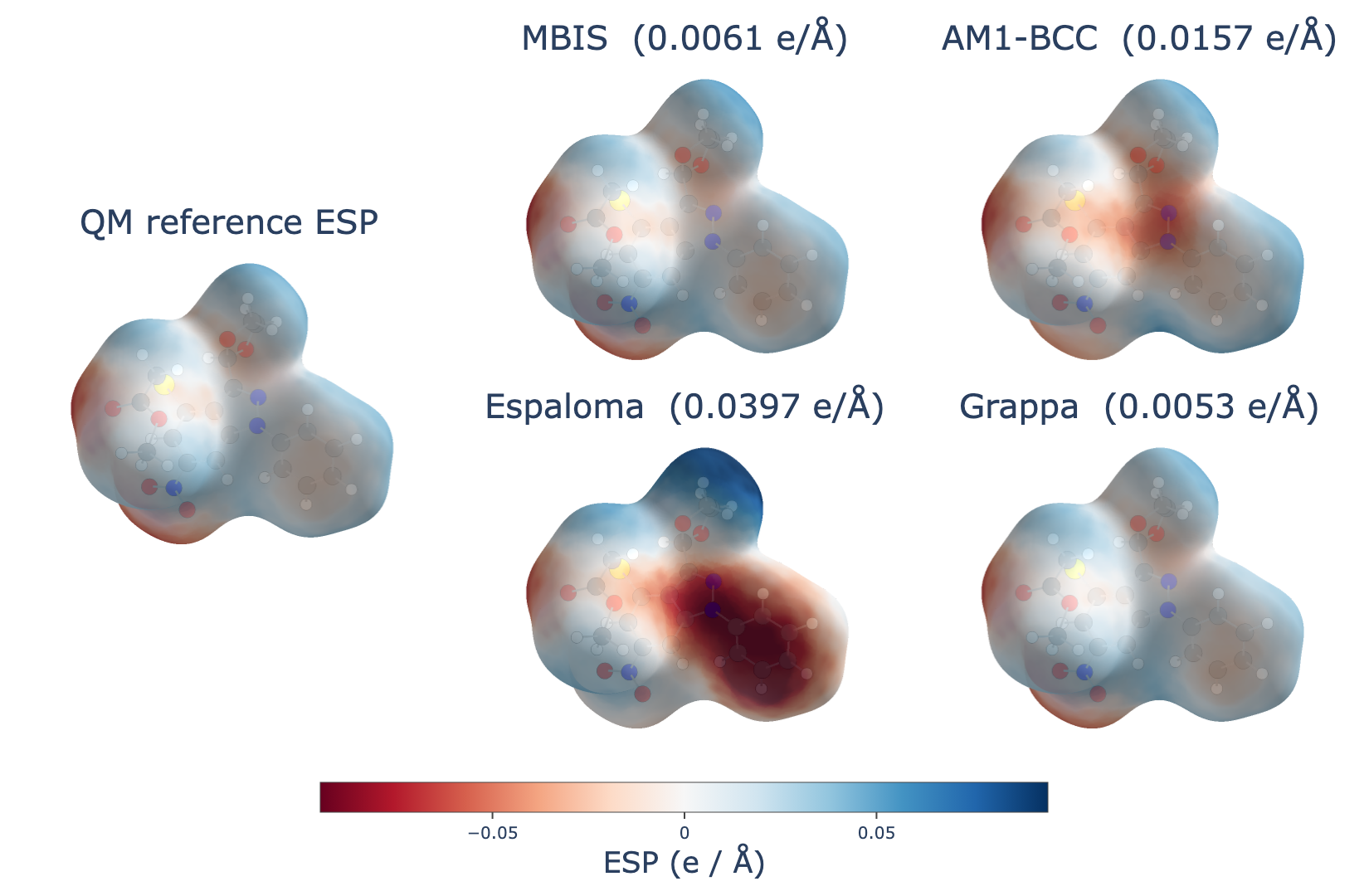}
    \caption{ESP mapped onto the $1.4\times$ vdW surface of a representative molecule.
    \gFF closely recovers the ESP surface derived from DFT with high accuracy, matching the computationally more expensive partial charges derived from the MBIS density partitioning. Standard empirical methods such as AM1-BCC and machine-learned charge models (espaloma-0.3.2) show noticeably higher deviations from the QM reference. 
    }
    \label{fig:esp_comp}
\end{figure}

\subsection{Cumulative Error Distributions for Geometry Optimization}
Table \ref{tab:force_field_comparison} in the main text reports the median and mean errors for the OpenFF Industry Benchmark. 
We complement these aggregate metrics with the full cumulative distribution functions (CDFs) for the Root Mean Square Error (RMSE), Torsion Fingerprint Deviation (TFD), and relative conformer energy errors ($|\text{ddE}|$) in Figure \ref{fig:geom_benchmark}, which provide a more detailed comparison. The distributions for \gFF and the bonded-only baseline \gBOND are nearly indistinguishable. This indicates that jointly learning the nonbonded parameters preserves the accuracy of hybrid ML/MM models that only predict bonded parameters. Thus, physical regularization constrains the additional learnable nonbonded degrees of freedom sufficiently so that they do not substantially alter the learned bonded interactions despite the underlying energy degeneracy.  
\begin{figure}[!h]
    \centering
    \includegraphics[width=\linewidth]{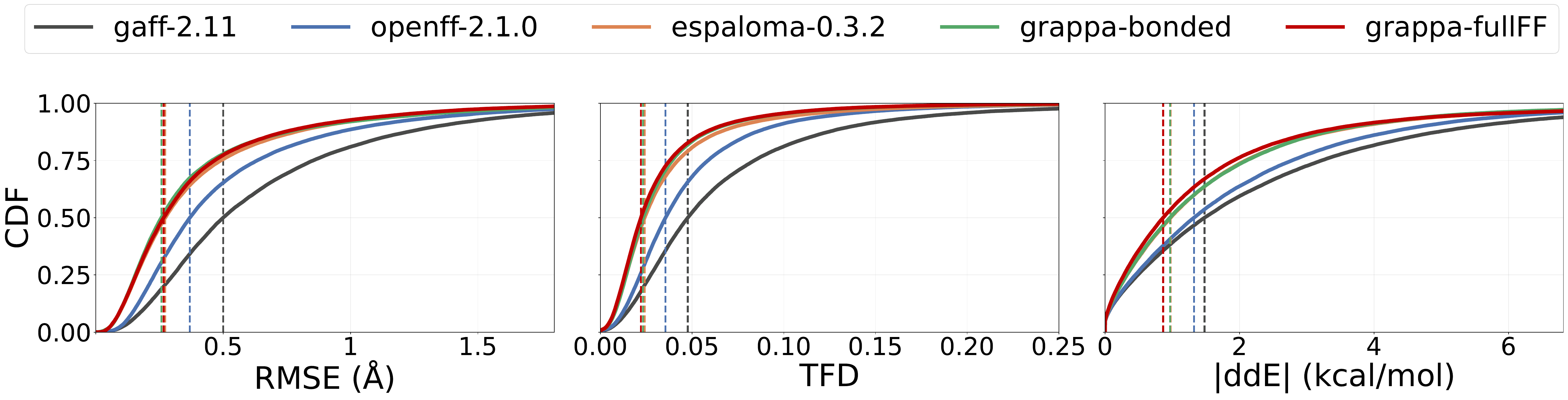}
    \caption{
    Cumulative distribution functions (CDFs) for the geometry Root Mean Square Error (RMSE), Torsion Fingerprint Deviation (TFD), and relative conformer energies (|ddE|) evaluated on the OpenFF Industry Benchmark Season 1 v1.1.
    The machine-learned force fields (espaloma-0.3.2,  \texttt{grappa-bonded}, \texttt{grappa-fullFF}) consistently outperform classical force fields (GAFF-2.11, OpenFF-2.1.0) on geometric fidelity.
    \gFF matches the structural accuracy of its bonded-only predecessor, confirming that the joint prediction of nonbonded parameters does not degrade bonded performance.
    }
    \label{fig:geom_benchmark}
\end{figure}

\subsection{Quantitative Divergence of Ramachandran Free-Energy Surfaces}
Although the Ramachandran FESs in Figure \ref{fig:md-fes} qualitatively show that the expected metastable states ($\alpha_R$, $\alpha_L$, $\beta$/PPII) are sampled, we provide quantitative metrics to evaluate the underlying thermodynamic distributions. To do so, we compute the 2D probability density distributions $P(\phi, \psi)$ from $500$ ns MD trajectories using a $5^\circ \times 5^\circ$ binning grid for each FF, and evaluate the symmetric KL and JS divergences relative to those obtained from AMBER99SB-ILDN. 

As shown in Table \ref{tab:rama_kl}, both \gBOND and \gFF models closely reproduce the AMBER distribution. \gFF achieves a JS divergence of just $0.0820$ nats, approximately twice as close to the reference distribution as \texttt{espaloma-0.3.2}. The difference between \gBOND and \gFF is negligible ($\Delta \text{JS} \approx 0.013$ nats), which quantitatively confirms that replacing established empirical nonbonded parameters (AMBER's TIP3P-compatible charges and LJ) with our jointly learned parameters does not change the complex conformational dynamics of the peptide in explicit solvent.

\begin{table}[!h]
\centering
\caption{Divergence of each force field's Ramachandran free-energy surface from AMBER99SB-ILDN, computed from a $5^{\circ}\!\times\!5^{\circ}$ histogram of $(\phi, \psi)$ over $500$~ns of unbiased TIP3P MD. Smaller is closer to AMBER99SB-ILDN.}
\label{tab:rama_kl}
\begin{tabularx}{\textwidth}{l *{2}{>{\raggedleft\arraybackslash}X}}
\toprule
Force field & Symmetric KL [nats] & Jensen--Shannon [nats] \\
\midrule
AMBER99SB-ILDN (reference)      & 0.000 & 0.0000 \\
espaloma-0.3.2                   & 1.752 & 0.1596 \\
\gBOND           & \textbf{0.432} & \textbf{0.0691} \\
\textbf{\gFF}  & 0.502 & 0.0820 \\
\bottomrule
\end{tabularx}
\end{table}


\section{Ablation of the physics-informed regularization}
\label{sec:app.ablation}
As discussed in Sec.~\ref{sec:training-and-regularize}, we use ESP to regularize the predicted charges rather than supervising them directly against reference charges. We compare this strategy against the alternative training objective in which predicted charges are supervised against MBIS reference charges through the MSE loss, 
\begin{align}
\label{eq:charge-loss}
\mathcal{L}_{q} = \frac{1}{N_\text{atoms}}\sum_i \left(\tilde{q}_i-q_i^\text{MBIS}\right)^2,
\end{align}
where $q_i^\text{MBIS}$ denotes the reference MBIS charge. This charge loss is incorporated as an additional term in Eq.~\eqref{eq:loss} with weight $\lambda_q$. 
Table~\ref{tab:ablation} lists four representative combinations of the hyperparameter sweep across $(\lambda_\text{ESP}, \lambda_q)$, all other regularization weights are fixed at $\lambda_E = 2$, $\lambda_\mathbf{F} = 1$, and $\lambda_v=100$. The last row corresponds to the configuration used to train \gFF.

Without either charge regularization, $\lambda_{\rm ESP}=\lambda_q=0$, the ESP RMSE increases by $35{-}56\%$ across all three data partitions compared to this work, and the ratio of mean charge magnitudes, $\langle q_i \rangle / \langle q_i^\text{MBIS} \rangle$, is only 0.74. Supervising charges alone ( $\lambda_q=1000$, $\lambda_\text{ESP}=0$) recovers the reference charge magnitude ($\langle q_i \rangle / \langle q_i^\text{MBIS} \rangle \sim 1$) but increases the ESP errors compared to our model. Adding ESP regularization alongside the MBIS charge loss further reduces the ESP error for PubChem and DES, but increases the error on the dipeptide set.
Using ESP regularization alone gives the lowest ESP errors for all three dataset partitions, even though the charge magnitude ratio degrades to $0.78$. These results indicate that even though ESP regularizations do not exactly recover MBIS charges, it can still provide a more accurate description of the electrostatics and electric response properties.

\begin{table}[!h]
\centering
\caption{Ablation of the two nonbonded regularization terms. $\lambda_q$ weighs the loss against per-atom MBIS charges as defined in Eq.~\eqref{eq:charge-loss}, $\lambda_\text{ESP}$ weighs the ESP loss defined in Eq.~\eqref{eq:esp_loss}. The energy, force, and volume-ratio loss weights are held fixed at  $(\lambda_E=2, \lambda_F=1,\lambda_v\!=\!100)$. $\langle q_i \rangle / \langle q_i^\text{MBIS} \rangle$ is the ratio of mean (absolute) charge magnitudes on 30 PubChem molecules. ddE MAE is from the geometry-optimization benchmark of Table~\ref{tab:force_field_comparison}.}
\label{tab:ablation}
\begin{tabular}{l c c c c c c c}
\toprule
Config & $\lambda_q$ & $\lambda_\text{ESP}$ & ESP$_\text{PubChem}$ & ESP$_\text{Dipep.}$ & ESP$_\text{DES}$ & $\dfrac{\langle q_i \rangle}{\langle q_i^\text{MBIS} \rangle}$ & ddE MAE \\
\midrule
No regularization    & 0    & 0    & 13.84 & 17.35 & 9.10 & 0.74 & --- \\
$\lambda_q \neq 0,\lambda_\text{ESP} = 0 $     & 1000 & 0    & 11.80 & 16.52 & 8.71 & 0.99 & 1.72 \\
$\lambda_q, \lambda_\text{ESP}\neq 0$         & 1000 & 100  & 11.07 & 20.24 & 7.55 & $\approx\!1.00$ & --- \\
$\lambda_q = 0, \lambda_\text{ESP}\neq 0$  & \textbf{0} & \textbf{1000} & \textbf{8.86} & \textbf{12.84} & \textbf{6.13} & \textbf{0.78} & \textbf{1.58} \\
\bottomrule
\multicolumn{8}{l}{\footnotesize ESP in units of $10^{-3} e/\text{\AA}$} \\
\end{tabular}
\end{table}

\section{Implicit Learning of Polar Hydrogen Radii}
\label{sec:app.H-TS-vdW}
By regularizing the LJ parameters toward TS-vdW volume ratios and simultaneously training on ab initio energies and forces, \gFF learns the environment-dependent LJ $\sigma$ values required to describe hydrogen bonding.
The TS-vdW volume ratios assign relatively large $\sigma$ values to polar hydrogens, whereas classical force fields assign smaller values to avoid unphysical Pauli repulsion in these hydrogen-bonded environments. As shown in Fig.~\ref{fig:grappa-predicted-sigmas}, \gFF learns this deviation from the TS-vdW reference directly from the energy and force gradients.

\begin{figure}[!h]
    \centering
    \includegraphics[width=1\linewidth]{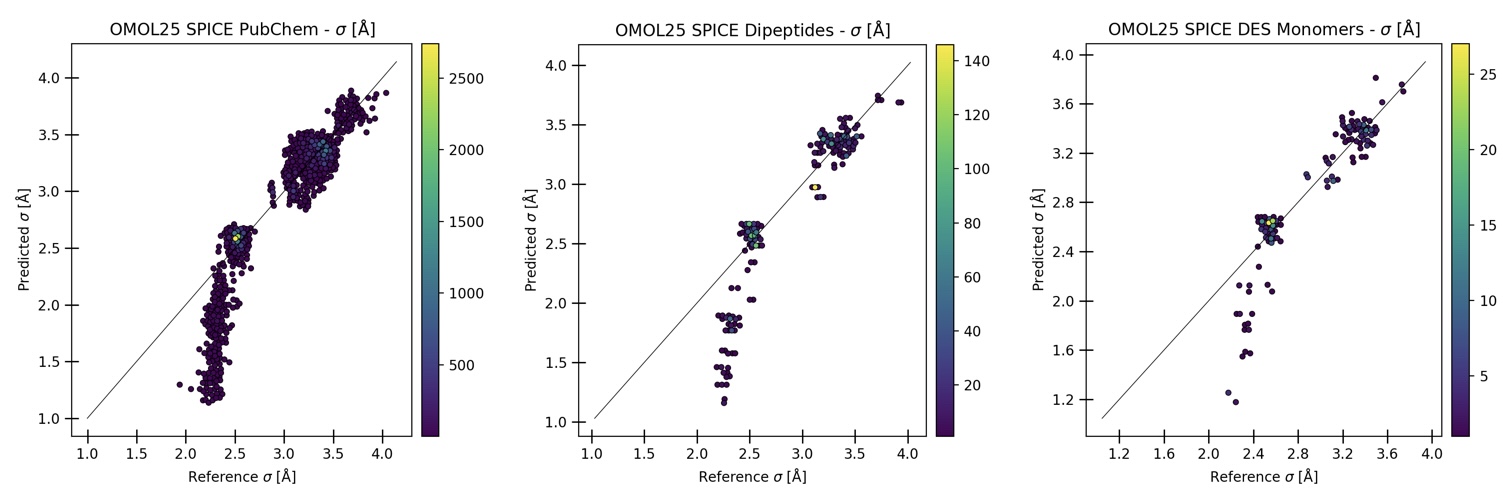}
    \caption{
    Comparison of predicted Lennard-Jones $\sigma$ parameters against the TS-vdW reference values for the three OMol25 SPICE subsets.
    The TS-vdW references assign relatively uniform radii to all hydrogen atoms ($\sigma\approx2.3-2.5\AA$), while \gFF learned a systematic downward spread ($\sigma\approx1.1-2.5\AA$). This correction is learned purely by training the model on ab initio energy and force gradients, which allows \gFF to accurately model hydrogen bonds without unphysical Pauli repulsion.
    }
    \label{fig:grappa-predicted-sigmas}
\end{figure}

\section{Molecular dynamics protocol}
\label{sec:app.md-protocol}
For the MD in Section \ref{sec:res.md} all four force fields, AMBER99SB-ILDN, \texttt{espaloma-0.3.2},  \texttt{grappa-bonded}, and  \texttt{grappa-fullFF}, simulate an identical system of a capped alanine dipeptide (ACE-ALA-NME) solvated in a cubic box of edge $2.4$~nm with $418$ TIP3P water molecules and one Na$^+$/two Cl$^-$ ions (0.15~M ionic strength) for $1{,}278$ total atoms.
The solvated topology and starting coordinates are generated once with OpenMM's \texttt{Modeller.addSolvent} and reused identically across all four force fields, so that any differences in the resulting Ramachandran FES (Fig.~\ref{fig:md-fes}) can be attributed to the force field rather than the simulated system. 

For the \gFF simulations, the solute (ACE-ALA-NME) was parameterized entirely by the NN, whereas the TIP3P water molecules were parameterized conventionally. This setup shows the stability of the learned nonbonded parameters at the solute-solvent interface. For \texttt{espaloma-0.3.2}, partial charges and bonded parameters are assigned to the initial solvated topology through a charge-assignment step, followed by equilibration and the simulation protocol described below.

Each system is relaxed, equilibrated for $100$~ps in the NVT ensemble followed by $1$~ns in NPT, and then propagated for $500$~ns in NPT at a $2$~fs timestep using OpenMM's \texttt{LangevinMiddleIntegrator} (friction coefficient $1\,\text{ps}^{-1}$) at $300$~K. The pressure is maintained at $1$~atm using a Monte Carlo barostat. Long-range electrostatics are treated with particle-mesh Ewald (real-space cutoff $1.0$~nm, default OpenMM Ewald error tolerance). Lennard-Jones interactions use the same $1.0$~nm cutoff with an isotropic long-range dispersion correction. 
All bonds to hydrogen are constrained, and water is kept rigid using OpenMM's default constraint solver. Frames from the resulting MD trajectory are saved every $1$~ps, yielding $500{,}000$ samples of $(\phi,\psi)$ per force field.



\end{document}